# The Influence of Policy Regimes on the Development and Social Implications of Privacy Enhancing Technologies

David J. Phillips, Assistant Professor
Department of Radio-Television-Film
University of Texas at Austin
djp@mail.utexas.edu

**ABSTRACT**

As privacy issues have gained social salience, entrepreneurs have begun to offer privacy enhancing technologies (PETs) and the U.S. has begun to enact privacy legislation. But "privacy" is an ambiguous notion. In the liberal tradition, it is an individualistic value protecting citizens from intrusion into a realm of autonomy. A feminist critique suggests that the social utility of privacy is to exclude certain issues from the public realm. Sociologists suggest that privacy is about identity management, while political economists suggest that the most salient privacy issue is the use of personal information to normalize and rationalize populations according to the needs of capital. While PETs have been developed for use by individual consumers, recently developers are focusing on the business to business market, where demand is stoked by the existence of new privacy regulations. These new laws tend to operationalize privacy in terms of "personally identifiable information." The new generation of PETs reflect and reify that definition. This, in turn, has implications for the everyday understandings of privacy and the constitution of identity and social life. In particular, this socio-technical practice may strengthen the ability of data holders to rationalize populations and create self-serving social categories. At the same time, they may permit individuals to negotiate these categories outside of panoptic vision. They may also encourage public discussion and awareness of these created social categories.

**TYPES OF PRIVACY**

"Privacy" is an ambiguous notion – different individuals and different social groups may entertain conflicting ideas of the utility of privacy and the danger of privacy invasion.

In the liberal tradition, privacy is an individualistic value. Privacy of this sort protects a sphere of autonomy in which the individual is free to be fully herself, and to enjoy intimate relations with others. It protects against intrusions both from the overbearing state and from the pressure of social norms. This type of privacy protects against Peeping Toms, warrantless searches, and laws restricting intimate behavior. It is perhaps characterized most aptly by Warren Brandeis, who defined privacy as the "right to be left alone" (Warren and Brandeis, 1890).

Some feminists have suggested that this particular privacy ideal does not benefit all individuals equally. Indeed, it is used in the construction of social inequality. As an ideal, it helps to create a private, domestic sphere apart from the public realm. To this private sphere are relegated certain populations, particularly women, and certain issues, particularly those of sexuality. The public/private distinction then serves as a tool of social silencing and repression. From this perspective, the most important privacy issues are not those of freedom from intrusion into the domestic realm, but instead of the social construction of the public/private divide itself. Feminist scholarship investigates the construction of that divide, and its implications for social power. Identity politics, particularly the public claiming of stigmatized identities, is very much concerned with this political construction of private realms (Boling 1996).

Another strand of privacy theory is also related to the construction of social identities. Sociologists, especially Goffman (1959), have explored how individuals selectively employ revelation and concealment to facilitate social performances. These performances create and maintain social identity as they re-inscribe social relations. Moreover, the performances are contextually specific. Certain roles are assumed in relation to certain others in certain places and circumstances. The ability to insist on a certain social performance is a measure of social power. Therefore social power includes the ability to control the informational context of a relationship. Waiters close the pantry door to withhold from the diners in the front stage knowledge of the workings of the kitchen (the back stage). Privacy – the ability to close the door - is implicated in this negotiation of social identities. This understanding of the use of privacy is perhaps best summed up in Westin's definition of privacy as an individual's right "to control, edit, mange, and delete information about them[selves] and decide when, how, and to what extent that information is communicated to others" (Westin 1967)

During the 1960's, new practices of computerized data collection re-enlivened privacy concerns. Originally these concerns were phrased in terms of the liberal individualist model of privacy, where threats of Big Brother loomed. Soon, however, some social theorists adopted Foucault's model of the Panopticon to understand privacy and surveillance practice. In this model, the focus is less on surveillance as it targets individuals, and more on surveillance as a technique of social organization which creates disciplines of knowledge, and normalizes and rationalizes populations in relation to those disciplines. In its idealized form, panoptic surveillance individualizes each member of the population, and permits the observation and recording of each individual's activities, then collates these individual observations across the population. From these conglomerated observations, statistical norms are produced relating to any of a multitude of characteristics. These norms are then applied back to the subjected individuals, who are categorized and perhaps acted upon, either with gratification or punishment, according to their relation to the produced norm. Thus surveillance produces both discipline (that is, conformity to the norm), and the disciplines (regulated fields of knowledge and expertise). The entire system gains efficiency when each individual realizes the inevitability of disciplinary action and internalizes the discipline, making the action itself unnecessary (Foucault 1979) . More recently, political economists have added that this process occurs overwhelmingly to serve the needs of capital, especially by normalizing and rationalizing consumer behavior in the marketplace (Gandy 1993b). This type of privacy theory addresses two major social issues. The first is the use of demographic and other consumer generated information to structure the lived world by creating categories of ideal consumers, markets, products and places. The second issue is the social discrimination inherent in the categorization and ranking of particular consumers by their potential value.

These types of privacy concerns, which we may, for the sake of convenience, label "freedom from intrusion", "public participation", "identity management" and "social coordination", are obviously interrelated. They are not offered as mutually exclusive truth claims as to the meaning of privacy, but as a sort of compass rose to the philosophical, cultural, and ideological terrain of privacy discourse.

Individuals and groups in different social positions may find that their interests in privacy vary in different contexts. For example, a gay man may attempt to make sexuality a public issue while simultaneously keeping private the specifics of his sexual behavior. He may seek political recognition of a gay community while resisting the construction of that community as a market segment. He may variously try to bring his sexuality to the fore, either as political fact or as erotic invitation, or suppress it as a matter of little moment.

## CONSUMER ORIENTED PETS AS MEDIATORS OF PRIVACY INTERESTS

Some social actors have negotiated their privacy concerns through the use of privacy enhancing technologies (PETs). These systems employ a number of different techniques and facilitate the protection of different types of privacy. Until recently, most of these systems have been designed to be adopted by the individual end user, rather than incorporated into an institutional setting.

Among the earliest PETs were cryptographic systems that protected communications between individuals from undesired eavesdropping. They ensured that messages sent to a particular individual could be read only by that individual. Pretty Good Privacy (PGP) is a prime example of this type of system (Garfinkel, 1995). These systems address the first type of privacy concern: freedom from intrusion.

Anonymizers were designed to break the link between a user's online interactions and the user herself. They make it difficult or impossible to trace the origin of a message. One of the earliest of these systems was pennet.fi. All e-mail to be anonymized went through a central server in Finland, which automatically stripped identifying information from the e-mail headers, and replaced it with more or less random identifiers. All mail seemed to originate at the anonymizing server, and only the server operator had access to the database that linked users' information to their anonymized identifiers. Other systems were established which performed similar operations on web traffic. This type of system can be seen to mediate the first and second types of privacy concern. They prevent intrusion since the physical space upon which to intrude remains unknown and so unreachable. More importantly, though, they permit the user to maintain a public presence and engage in public interactions without fear of retribution. By altering the established mechanisms of social silencing, these systems allow users to bring taboo topics into the light. In that sense, they alter the means of constructing the public/private divide.

Later, systems were developed whose aim was to control the unintentional flow of information from an individual to a corporate correspondent. The design of standard browsers permits web sites to place "cookies" on a user's machine and read those cookies during subsequent visits to the site. Cookies are often used to uniquely identify an individual user's machine. The first time a machine is used to access a particular site, that site sets a unique id cookie on the machine, and also establishes a record indexed by that identifier in an on-site database. Thereafter, whenever the machine is used to access the site, that site reads the unique identifier in the cookie, monitors the user's behavior at the site, and updates the record accordingly. In this way, the site operator (or, more often, third party profilers) create comprehensive profiles of each user's web browsing habits. PETs which interrupt this process include pseudonymizers and cookie management software.

Cookie management software permits the user to obliterate cookies and other persistent identifiers and so establish a new digital identity with each browser session. This interferes with the ability to collect information across browser sessions, and to correlate that information into a single record to be used as an input to statistical processes. It therefore is informed by the fourth privacy concern – panoptic social coordination.

Pseudonymizers permit the user to establish a number of separate, but persistent, identifiers among which she may choose for her web interactions. Each pseudonym (or simply "nym") would have a separate "cookie jar" to which web sites could write and read. Therefore, each nym would eventually be represented by separate, unlinkable, digital records. Freedom®, from Zero Knowledge Systems, was, until its recent demise, the most technically sophisticated of pseudonymity suites (McFarlane et al, 2000). In addition to acting as a pseudonymizer, it was also an effective anonymizer, in that it made it very difficult to establish linkages between a user and that user's pseudonymous online presence. These types of systems are primarily informed by the third type of privacy - context and identity management. However, in addressing anonymity, they also address privacy through the frames of intrusion and participation.

Another type of PET does not so much allow the user to establish different identities as it permits the user to negotiate the revelation of sets of data points of a single personae. These are infomediaries, like Lumeria or Privaseek. With these, the user establishes a set of rules under which she is willing to divulge certain types of information about herself. If an information seeker signals that he is willing to abide by certain constraints (perhaps offering the user a discount, or promising to restrict third party access to the information) then the infomediary will release the information. This type of privacy protection is informed primarily by the third privacy ideal – context and identity management.[1] It presumes an autonomous individual agent with relatively fixed identity, yet it recognizes that certain information pertaining to that agent is sensitive in some contexts but not in others. By incorporating the ability to bargain for the right to access personal data, it also recognizes the fourth understanding of privacy issues – that is, the role of personal information as a valuable input into the construction of market relations.

These systems have not been widely adopted by individual internet users for at least four reasons. First, they are often operationally complex. PGP requires an infrastructure for distributing cryptographic keys which users are likely to find impenetrable (Whitten and Tygar, 1999). The cryptographic problem of establishing unlinkable, but persistent and viable pseudonyms through open Internet protocols is extraordinarily difficult. This functional complexity makes it difficult to design an interface which is anything but inscrutable, slow, and erratic. Second, and even presuming a robust and transparent interface, it is not at all clear what one is to *do* with numerous persona in online interactions. The management of face and identity is second nature in face to face interactions. We do not usually, for example, present ourselves at work as we do at a college football game. But these identity producing activities – modes of dress or speech – are more or less taken for granted and culturally commonplace. Slippage between identities is almost unconscious. In online identity management, the choice of identities is very much a conscious one.  But it is a decision made outside of familiar physical and social contexts.  It is not clear to the user just who is likely to be present in an online context, what the appropriate norms of behavior are, and what the consequences are for transgressing those norms. In short, it is not clear why or to what end one would choose one persona over another in any online context (Phillips, 2001).  Third, there is the problem of social and economic power. This is particularly an issue with infomediary systems, which are founded conceptually on the idea of a market negotiation for bits of personal data.  But in a market comprised of individuals in negotiation with corporations, individuals are likely to be contract takers.  This is particularly true in a market for personal information, where the marginal utility to the individual of one piece of data is small enough to make its defense economically unfeasible, while the organization has every economic reason and resource to protect and expand its data collection (Gandy 1993a). Finally, the frames though which privacy issues are presented, both in policy debate and in the popular press, overwhelmingly privilege the understanding of privacy as a personal, rather than a social, value (Regan, 1995; Phillips and Curry, forthcoming). When the issue is framed as a balancing of the rights of the individual versus the needs of society, it is difficult to justify the social or personal effort to protect personal information.

---

[1] Indeed, Lumeria's public relation's material is quite explicit in this regard:
"But, what is privacy? In the Internet age, the definition of privacy provided by Alan Westin in Privacy and Freedom seems most fitting: '…the claim of individuals, groups, or institutions to determine for themselves when, how, and to what extent information about them is communicated to others.' Whereas privacy issues used to apply to what Chief Justice Brandeis called 'the right to be left alone,' Lumeria believes that privacy is not about hiding from others, but rather about controlling the flow of your personal data." ("What is Privacy?")

### B-2-B PETS

Although consumer adoption of individual-to-individual and consumer-to-business systems has been sparse, activist organizations have successfully pressed for the passage of privacy laws and regulations. At the same time, privacy entrepreneurs have attempted to stimulate institutional adoption of PETs. One of these attempts is Zero Knowledge System's Enterprise Privacy Rights Management (PRM) Suite. This suite is aimed at database holders, rather than consumers, and is designed to mediate the exchange of data both among large data holders and between individuals and data holders. It is not fully deployed yet, so what follows is a description of the apparent intent, rather than the implementation of this system.

PRM is designed to allow companies to "tap their databases ... without violating privacy laws" and to "create new offers to customers, without violating privacy." (Riga, 2001) It enables "management of personal information in a manner that supports business objectives, customer preference and choice, and the global requirements of privacy regulations" through "the application of rules based on regulation, corporate policy, and/or customer preferences to personal information" ("The Privacy Challenge…", "Privacy Rights Management ... ").

Note that the system is designed to incorporate regulation as a subset of its privacy structure, while also incorporating internal policy and consumer choice. However, compliance with policy is likely to be, not only the least common denominator, but the full extent of its protection. This is, again, because individuals haven't the wherewithal to knowledgeably adopt and use PETs, nor the market information or market power to strike self-benefiting contracts, nor the ideology to understand privacy as a social good, rather than an individual benefit.

Before we can determine the likely implications of these systems, then, we must examine the regulations they are likely to mediate.

## PRIVACY REGULATION AND IDENTIFICATION

There are three major pieces of privacy legislation in the U.S. These are the Children's Online Privacy Protection Act (COPPA), parts 160 and 164 of the Health Insurance Portability and Accountability Act (HIPAA), and Sections 6801 through 6810 of the Gramm-Leach-Bliley Act (GLB). All of these operationalize privacy concerns by restricting the collection, use, or transfer of "personally identifiable information". As we shall see, the operationalization of privacy in terms of a particular kind of identification practice has implications for the kinds of social relations facilitated by these regulatory regimes.

What does it mean for information to be "personally identifiable"? "Identification," like "privacy," is an overloaded word. In general, we may identify three different kinds of identification practice. *Lexical identification* links a name to an entity. Anyone who has spent much time with horticulturalists has enjoyed this type of activity, as autumn walks become an opportunity to identify, in Latin, every rooted thing. *Indexical identification* points to an entity. Yelling "Stop! Thief!" is an act of indexical identification. It is a link to a particular entity in a particular time and place. *Descriptive identification* assigns attributes to an entity in a way that places it in relation to other entities. To refer to someone as a typical disgruntled academic is descriptive identification (Agre, 1999).

As with our types of privacy, these are not mutually exclusive identification practices. Botanical names are both lexical and descriptive, and "unique identifiers" such as the social security number are lexical on their face, but deeply embedded in practices of indexical identification. They help to track and find you.

Identification practices are also implicated in the type of privacy practices discussed earlier. Descriptive and lexical identification are used in the social processes of statistical production. That is, it is essential in statistics to know the particular attributes of an entity. It is also necessary to know whether or not an entity with the same attributes is the same entity, or another occurrence of the same *type* of entity. Descriptive and lexical identification are implicated, then, in those parts of the surveillance process which create social norms and fields of knowledge. Indexical identification, the actual pointing to or accessing of a particular bodily entity, is implicated in that part of the surveillance process which applies those produced norms to individuals. Because indexical identification is an element of panoptic normalization, it is part of the fourth privacy concern – social coordination. Because it permits action upon a particular individual, it is cognitively linked to the first privacy concern – intrusion.

The three laws mentioned above are concerned only with indexical identification, not with descriptive or lexical identification. That is, they consider within the realm of privacy protection only information potentially *linked* to an individual's body, not information *produced* by an individual, *pertaining* to an individual, or *describing* an individual.[2]

COPPA is most explicit in this regard. It prohibits the collection or dissemination of personal information of children under the age of 13, and defines "personal information" to mean "individually identifiable information", including names, addresses, telephone numbers or "any other identifier that ... permits the physical or online contacting of a specific individual." The collection of other information, for example age, patterns of

---

[2] I am not a lawyer, nor do I play one on TV. These are recent laws, and have not yet been subject to any interpretive rulings. The discussion that follows is informed only by my own face-value reading of the texts.

web usage, or preferences in breakfast cereals, is prohibited only if it is combined with such an identifier. (15 USCS § 6501 (2001))

GLB, too, restricts the dissemination only of "nonpublic personal information". This is defined as "personally identifiable financial information" provided by a customer in any transaction with a financial institution. The restriction on dissemination includes "any list, description, or other grouping of consumers (and publicly available information pertaining to them) that is derived using any nonpublic personal information other than publicly available information." (15 USC, Subchapter I, Sec. 6809 (4))

HIPAA incorporates an extremely complex and protective set of rules regarding the dissemination of "identified information." However, "de-identified information" is specifically excluded from the rules. Records are considered de-identified if the following fields are removed: Names; geographic subdivisions smaller than a State, dates of birth, admission, discharge, treatment, or death (except for the year of such dates), telephone or fax numbers, social security numbers, e-mail addresses, URLs, IP addresses, medical record, health plan, and other account, license, serial, or other unique identifying numbers, and biometric identifiers (including facial photographs). Importantly, however, a record holder may assign a code to a de-identified record in order to permit the original record holder to re-identify the record. (45 CFR 164.514(b)(2)(a), 45 CFR 164.514(c), 65 Fed. Reg. at 82818)[3]

In summary, these rules permit data holders to do anything they want with information derived from and produced by individuals, provided that they de-link the data from the individual's *indexical* identity. The actions of individuals may be used to describe and model them in any way the data holder wishes, provided the data holder cannot actually contact, locate, point to, or act upon the individual thus described. They are, again, informed by an understanding of "privacy" that is closely related to the protection of the individual from unwanted intrusion.

---

[3] For a discussion of the problems inherent in de-identifying records, see Sweeney 1997.

However, it is technically and administratively possible within this legal framework to produce profiles which can be shown to pertain to a specific, unique individual, and yet are "de-identified" in the sense that no one can point to any particular person and say "this is about him." That is, profiles can be identified lexically but not indexically. This can be accomplished through one way hash functions. A one way function is a function which relatively simple to compute, but extremely difficult to reverse. A one-way hash function accepts a data string as input and produces as output a seemingly random data string, the hash value. The same input will always produce the same hash value. However, while it is simple to compute the hash value of a data string, it is virtually impossible to compute the original data string given only its hash value. (Schneier, 1996)

Therefore, a data holder may code a record by computing a hash value from an identification field in the record, say the social security number, and attaching that hash value to the record. The holder then removes all identifying fields from the record. There is now no way for the record holder to reverse the hash, recover the social security number, and re-identify the record. Legally, that record is not subject to any privacy regulation. However, another data holder may de-identify their records using the same algorithm. If they use the same hash fields and functions, then the records pertaining to the same individual share the same hash value, while records pertaining to different individuals will have different hash values. The records will be lexically and descriptively identified, but not indexically identified.

This technique, born of a particular understanding of "privacy" as articulated in regulation and mediated through computing power, has significant implications for the social ordering of identity and categories.

**SOCIAL IMPLICATIONS**

This potential new informational structure breaks the link, not between the different records created by the actions of a single individual, but between those records and the fleshly individual herself. Philosophically and practically, these practices facilitate a world that consists of unary individuals (that is, unique individuals whose identity is constant across contexts), who are constituted as data objects by invisible and powerful observers. Thus the part of the panoptic model that permits the production of knowledge, of disciplines and statistical norms remains intact. Data holders will still be in the powerful position of creating and imposing an ontology of the world. Data produced by individuals will still be used to construct pricing schemes, risk management schemes, persuasive ads, and appropriate places. Seen through the lens of the fourth vision of privacy, which concerns itself with the rationalization of population, this data protection scheme at first seems to offer no protection at all.

However, the ability for data holders and marketers to actually categorize and act upon people in light of these constructed orderings may be restricted. That is, though marketers may be able to develop an ideal consumer and product developers may be able to fashion an ideal product for that person, they may yet be unable to tell whether, or to what extent, any particular individual fits that ideal. Individuals may be more able to choose their presentation to fit their context. The third sense of privacy, identity management, may be strengthened and the processes of surveillant social organization may be disrupted.

Such a system also addresses the second sense of privacy – that of public participation in the definition of the public/private divide. One of the most insidious aspects of market profiling is that the social models thus produced are private property. For example, the algorithms by which auto insurance companies determine the rate for each individual policy holder are trade secrets. Although the individual has a right to know what data was used to calculate the rate, they will not be told how that rate was calculated, or what they may do to reduce the rate. Not only are the models trade secrets, but marketers, aware of the public's increasing concern for privacy, make specific public relations efforts to disguise the extent of their knowledge. When this private social modeling is combined with the private persuasive techniques of targeted marketing, the result is an anti-democratic, anti-deliberative process of social shaping. While the effects of mortgage and insurance redlining are enduring and evident, the effects of price discrimination through targeted marketing are socially invisible. Once the ability to target individuals for persuasive campaigns is reduced, these campaigns must again take place in the public sphere. Therefore these laws and informational systems also address the second privacy problem – the strategic maintenance of the public/private divide.

It is important to recognize that these predictions are merely that. The interactions among systems developers, laws, corporate data holders, activist organizations, news media, and individual data providers is extremely complex and subject to the vagaries of each historical moment. What I have laid out is merely one set of pressures, one pattern of mutual influence in an all but unfathomable set of social relations.

**CONCLUSION AND RECOMENDATIONS**

In fashioning COPPA, HIPPA, and Gramm-Leach-Bliley, U.S. legislators have employed the cognitive frame of personal intrusion to understand privacy concerns. They have ignored other possible understandings, most obviously that of social coordination and control. They have structured a market for personal data such that the originator of the data, and the subject of the data, have rights over the data only in certain circumstances – that is, only when the data can be used to physically and individually affect the data subject. In doing so, they have laid the regulatory foundation for information systems which substantially leave intact the ability of large data holders to create systems of social knowledge and power. They also, however, marginally limit their ability to impose those knowledge systems on the surveilled population. On the whole though, they constitute a re-entrenchment, though also a re-alignment, of surveillance power.

Should legislators wish to create a more fundamentally equitable information regime, they should consider operationalizing data laws not in terms of information that identifies individuals, but instead information that is produced by or pertains to individuals, whether those individuals are identifiable or not. For example, Canada's Personal Information Protection and Electronic Documents Act (PIPEDA) is a small step in this direction. Although again referring only to ''personal information,'' the Act prohibits not only the transfer of that information, but its use as well. Prohibitions on "use" might well include the de-identifying of those records in preparation for their transfer. More importantly though, activists should encourage legislators to see privacy issues outside of the frame of personal intrusion, and more through the frame of social organization. Then legislation might be founded on a moral right, rather than a property right or a privacy right, to the use of the traces of one's life in structuring the world in which we all live.

# REFERENCES

Agre, Philip. 1999. "The architecture of identity: Embedding privacy in market institutions" Information, Communication and Society 2(1): 1-25.

Boling, Patricia. 1996. Privacy and the Politics of Intimate Life. Cornell University Press: Ithaca, NY.

Foucault, Michel. 1979. Discipline and Punish. (tr. Alan Sheridan) New York: Vintage Books.

Gandy, Oscar. 1993a. Toward a Political Economy of Personal Information. Critical Studies in Mass Communication 10: 70-97.

Gandy, Oscar. 1993b. The Panoptic Sort: A Political Economy of Personal Information. Boulder, CO: Westview Press.

Garfinkel, Simson. 1995. PGP: Pretty Good Privacy. Sebastopol, CA: O'Reilly and Associates.

Goffman, Erving. 1959. The Presentation of Self in Everyday Life. New York, NY: Doubleday.

Lyon, David. 1994. The Electronic Eye: The Rise of Surveillance Society. Minneapolis: University of Minnesota Press.

McFarlane, Roger, Adam Back, Graydon Hoare, Serge Chevarie-Pelletier, Bill Heelan, Christian Paquin, and Deniz Sarikaya. 2000. Freedom® 2.0 Mail System. Montreal, CA: Zero-Knowledge Systems Inc.

Phillips, David J. 2001. "Freedom® and Privacy: the Role of Privacy Enhancing Technologies in Structuring Social Identity." Annual Conference of the Society for Social Studies of Science. November 1-4, 2001. Cambridge, MA.

Phillips, David J. and Michael Curry. (forthcoming). "Geodemographic Practice, Privacy, and the Changing Relation of People and Place." In Lyon, David (Ed.) Surveillance as Social Sorting: Privacy, Risk, and Automated Discrimination. London: Routledge.

"Privacy Rights Management Technology"
http://www.zeroknowledge.com/business/privacyrights.asp; accessed 22 Sep 01

"The Privacy Challenge for Business"
http://www.zeroknowledge.com/business/default.asp; accessed 22 Sep 01

Regan, Priscilla M. 1995. Legislating Privacy : Technology, Social Values, and Public Policy. Chapel Hill: University of North Carolina Press.

Riga, Andy. 2001. "Prophets zero in on profits." The Gazette (Montreal):(Mar 28); P. D1

Schneier, Bruce. 1996. Applied Cryptography, Second Edition. Wiley: New York, NY.

Sweeney, Latanya. 1997. "Weaving Technology and Policy Together to Maintain Confidentiality" Journal of Law, Medicine, and Ethics 25: 98-110.

Warren, Samuel and Louis Brandeis. 1890. "The Right to Privacy" Harvard Law Review 4: 193.

Westin, Alan F. 1967. Privacy and Freedom. New York: Atheneum.

"What is Privacy?" Undated. Lumeria, Inc. http://www.lumeria.com/paper1/1.sthml. Accessed 18 September 2001.

Whitten Alma, and J. D. Tygar. 1999. "Why Johnny Can't Encrypt: A Usability Evaluation of PGP 5.0". The 8th USENIX Security Symposium. August 23 – 26, 1999. Washington, D.C. http://www.usenix.org/publications/library/proceedings/sec99/full_papers/whitten/whitten_html/. Accessed 16 October 2001